\def\mycmd{2}
\if\mycmd1
\documentclass[11pt,onecolumn, draftcls]{IEEEtran}
\else 
\documentclass[10pt,a4paper,twoside,journal]{IEEEtran}
\fi
\usepackage{tabularx}

\usepackage{multicol}
\usepackage{multirow}
\usepackage{float}
\usepackage{mathtools}
\usepackage{amsmath}
\usepackage{amsthm}
\usepackage{amssymb}
\usepackage{graphicx}
\usepackage{transparent}
\usepackage{wasysym}
\usepackage{setspace}
\usepackage{color}
\usepackage{bm}
\usepackage{cases}
\usepackage{booktabs}
\usepackage{adjustbox}
\usepackage{subfigure}
\usepackage{glossaries}

\if\mycmd1
\usepackage[top=1.6cm, left=1.6cm, bottom=1.6cm, right=1.6cm]{geometry}
\else
\usepackage[top=1.4cm, left=1.4cm, bottom=1.4cm, right=1.4cm]{geometry}
\fi
\usepackage{tikz}

\usepackage{amsfonts}
\usepackage{makecell}
\usepackage{pbox}
\usepackage{epsfig}

\makeatletter

\floatstyle{ruled}
\newfloat{algorithm}{tbp}{loa}
\providecommand{\algorithmname}{Algorithm}
\floatname{algorithm}{\protect\algorithmname}

\theoremstyle{plain}

\theoremstyle{plain}

\theoremstyle{plain}

\usepackage{epsfig}
\usepackage{cite}
\usepackage{stfloats}
\usepackage{graphicx}
\usepackage{array}
\usepackage{enumerate}
\usepackage{enumitem}
\setlist[itemize]{leftmargin=*}
\setlist[enumerate]{leftmargin=*, label=\arabic*)}
\usepackage{graphicx}
\usepackage{times}
\usepackage{algorithm}
\usepackage{algpseudocode}
\theoremstyle{remark}

\makeatother

\algrenewcommand\algorithmicindent{1.0em}%
\providecommand{\lemmaname}{Lemma}
\providecommand{\propositionname}{Proposition}

\providecommand{\theoremname}{Theorem}
\providecommand{\theoremname}{Definition}
\newcommand{\rom}[1]{\uppercase\expandafter{\romannumeral #1\relax}}

\DeclarePairedDelimiter\floor{\lfloor}{\rfloor}

\algnewcommand{\IIf}[1]{\State\algorithmicif\ #1\ \algorithmicthen}
\algnewcommand{\ElseIIf}[1]{\algorithmicelse\ #1} % <<< added
\algnewcommand{\EndIIf}{\unskip\ \algorithmicend\ \algorithmicif}

\newcounter{problem}
\newcounter{save@equation}
\newcounter{save@problem}


\numberwithin{save@problem}{subsection}
\numberwithin{save@equation}{subsection}

\begin{document}
\title{Joint Optimization on Uplink OFDMA and MU-MIMO for IEEE 802.11ax: Deep Hierarchical Reinforcement Learning Approach}
\author{Hyeonho Noh,~\IEEEmembership{Student Member,~IEEE}, Harim Lee, and Hyun Jong Yang,~\IEEEmembership{Member,~IEEE}
\thanks{Hyeonho Noh and Hyun Jong Yang are with Department of Electrical Engineering, Pohang University of Science and Technology (POSTECH), Pohang, 37673, Korea (e-mail: \{hyeonho, hyunyang\}@postech.ac.kr).
Harim Lee is with School of Electronic Engineering, Kumoh National Institute of Technology, Gumi, Gyungbuk, 39248, Korea (e-mail: hrlee@kumoh.ac.kr).
Hyun Jong Yang is the corresponding author.}
}

\maketitle
\begin{abstract}\label{abstract}
This letter tackles a joint user scheduling, frequency resource allocation (USRA), multi-input-multi-output mode selection (MIMO MS) between single-user MIMO and multi-user (MU) MIMO, and MU-MIMO user selection problem, integrating uplink orthogonal frequency division multiple access (OFDMA) in IEEE 802.11ax. Specifically, we focus on \textit{unsaturated traffic conditions} where users' data demands fluctuate.
% Unlike typical cellular communication systems with fixed data frame lengths, in 802.11ax, the frame length is an additional optimization variable, more complicating the joint USRA and MIMO MS problem, especially given the standard's limited frequency band use cases. 
In unsaturated traffic conditions, considering packet volumes per user introduces a combinatorial problem, requiring the simultaneous optimization of MU-MIMO user selection and RA along the time-frequency-space axis. Consequently, dealing with the combinatorial nature of this problem, characterized by a large cardinality of unknown variables, poses a challenge that conventional optimization methods find nearly impossible to address. 
In response, this letter proposes an approach with deep hierarchical reinforcement learning (DHRL) to solve the joint problem. Rather than simply adopting off-the-shelf DHRL, we \textit{tailor} the DHRL to the joint USRA and MS problem, thereby significantly improving the convergence speed and throughput. 

\end{abstract}


\newacronym{USRA}{USRA}{user scheduling and resource allocation}
\newacronym{UL MU-MIMO}{UL MU-MIMO}{uplink multi-user multi-input multi-output}
\newglossaryentry{MU-MIMO}{
name=MU-MIMO,
description={a},
parent=UL MU-MIMO
}
\newglossaryentry{MIMO}{
name=MIMO,
description={a},
parent=UL MU-MIMO
}

\section{Introduction}
\label{sec:introduction}
IEEE 802.11ax, as the inaugural standard capable of concurrently supporting \gls{UL MU-MIMO} and orthogonal frequency division multiple access (OFDMA), has been pivotal since its release \cite{IEEEstd21}. Evidently, tackling the joint \gls{USRA} problem that encompasses elements such as \gls{UL MU-MIMO}, OFDMA, \gls{MU-MIMO} user selection, and MIMO mode selection (MS) between single-user (SU)-MIMO and MU-MIMO is crucial to optimize the system performance \cite{Wang18, Sangdeh21}. 
{In particular, the majority of studies \cite{Wang18, Sangdeh21, bankov18, Ha21, Lee19} tackling the joint problem have evaluated theoretical spectral efficiency under saturated traffic scenarios, assuming all users possess unlimited packet queues of infinite length. This assumption, however, diverges from real-world conditions, where user traffic is usually unsaturated, i.e., packet volumes can vary across different users, and packet length is subject to stringent constraints. 
Thus, it becomes imperative to address unsaturated traffic conditions to closely reflect real-world scenarios \cite{Xie19, Liu23, Bhattarai19}.}

Despite its significance, only a few traffic-aware \gls{USRA} studies have been researched in the realm of Wi-Fi and 3GPP-based cellular networks. 
However, these studies avoid inherent explosive complexity by neglecting one or several aspects of the joint problem, as summarized in Table \ref{tab:compare}. 
Even studies employing deep learning or reinforcement learning (RL), which have been widely employed in addressing complex problems under dynamic natures, have failed to fully account for all aspects of the joint problem.
Consequently, these oversights result in notable losses in generality and performance in the existing works.
Furthermore, the seamless adaptation of \gls{USRA} techniques designed for 3GPP-based cellular networks to 802.11ax proves challenging due to the constraints on frequency band utilization in 802.11ax \cite{Sangdeh21, bankov18}.

\begin{table}
\centering
    \caption{Comparison of the proposed scheme to previous works.
    }
    \vspace{-7pt}
    \adjustbox{width= \if 1\mycmd 0.6 \else 0.85 \fi \columnwidth}{
    \begin{tabular}{c|c|c|c|c|c|c|c|c}
    \toprule
    & \textbf{Ref} & \textbf{US} & \textbf{RA} & \textbf{OFDMA} & \textbf{MIMO MS} & \textbf{Buffer} & \textbf{MU-MIMO} & {\textbf{DL/RL}} \\ \midrule[\heavyrulewidth]\midrule[\heavyrulewidth]
    \multirow{2}{*}{Celluar} & \cite{Xie19} & \texttt{O} & \texttt{X} & \texttt{X} & \texttt{X} & \texttt{O} & \texttt{O} & \texttt{X} \\ \cline{2-9}
    & \cite{Liu23} & \texttt{O} & \texttt{O} & \texttt{O} & \texttt{X} & \texttt{O} & \texttt{O} & \texttt{X} \\
    \cline{1-9}
    % \multirow{5}{*}{Wi-Fi} & \cite{bankov18} & \texttt{O} & \texttt{O} & \texttt{O} & \texttt{X} & \texttt{X} & \texttt{X} & \texttt{X} \\\cline{2-9}
    \multirow{5}{*}{Wi-Fi} & \cite{Ha21,Lee19} & \texttt{O} & \texttt{O} & \texttt{O} & \texttt{X} & \texttt{X} & \texttt{O} & \texttt{X} \\\cline{2-9}
    & \cite{Wang18} & \texttt{O} & \texttt{O} & \texttt{O} & \texttt{O} & \texttt{X} & \texttt{O} & \texttt{X} \\ \cline{2-9}
    & \cite{Sangdeh21} & \texttt{O} & \texttt{O} & \texttt{O} & \texttt{O} & \texttt{X} & \texttt{O} & \texttt{O} \\ \cline{2-9}
    & {\cite{Balakrishnan19, Kotagiri21}} & \texttt{O} & \texttt{O} & \texttt{O} & \texttt{X} & \texttt{O} & \texttt{X} & \texttt{O} \\ \cline{2-9}
    & \cite{Bhattarai19} & \texttt{X} & \texttt{O} & \texttt{O} & \texttt{X} & \texttt{O} & \texttt{X} & \texttt{X} \\ \midrule
    % \midrule[\heavyrulewidth]
    \textbf{Wi-Fi} & \textbf{Ours} & \textbf{\texttt{O}} & \textbf{\texttt{O}} & \textbf{\texttt{O}} & \textbf{\texttt{O}} & \textbf{\texttt{O}} & \textbf{\texttt{O}} & \textbf{\texttt{O}} \\ 
    \bottomrule
    % \\
    \end{tabular}
    }
    \label{tab:compare}
    \vspace{-16pt}
\end{table}

% \begin{figure}[t]
%     \centering
%     \includegraphics[draft=false, width= \if 1\mycmd 0.5 \else 0.7 \fi \columnwidth]{Figures/conv_proposed.pdf}
%     \vspace{-7pt}
%     \caption{The comparison between the conventional and proposed DHRL model in a 20 MHz channel.}
%     \label{fig:off_the_shelf}
%     \vspace{-15pt}
% \end{figure}

% Deep learning, particularly deep reinforcement learning (RL), has been widely employed for solving NP-hard problems like \gls{USRA} in wireless communication, where considerations of dynamic states such as time-varying channel state information (CSI) and packet volumes of individual stations (STAs) are imperative \cite{Aung23, Luong19}.
% Nevertheless, when it comes to complex joint problems, a deep RL agent may struggle to capture the relationships among the problems. 
% To this end, deep hierarchical RL (DHRL) has been proposed \cite{Kulkarni16, Frans18}, decomposing a joint problem into a hierarchy of sub-problems, each with its own agent. 
% The top layer defines the overall goal of the problem, while the lower layers define the primitive actions necessary to maximize the overall reward.

In this letter, we address the \gls{USRA} optimization problem which involves UL MU-MIMO, OFDMA, MU-MIMO user selection, and MIMO MS for IEEE 802.11ax uplink under unsaturated traffic conditions.
{Specifically, we tailor a deep hierarchical RL (DHRL) framework, specialized for solving complex joint problems\cite{Kulkarni16, Frans18}, to the joint problem. After decomposing the \gls{USRA} problem into a hierarchy of RA and US, we define master and sub-agents for solving RA and US problems, respectively.
The proposed DHRL model can adapt to the dynamic wireless environment, enhance the sample efficiency, and improve network performance significantly. Specifically, unlike existing works, we integrate all aspects of the joint problem without any relaxation, ensuring our approach does not incur any loss in generality or performance.}

While applying off-the-shelf DHRL directly makes the learning process nearly impractical due to agents' explosive action space size, we refine sub-agents to align with the standard-defined frequency band utilization, markedly improving the practicality and performance of the system.
In addition, we propose a state update process based on channels' semiorthogonality, which leads to a huge performance gain.

{The rest of this letter is organized as follows: Section \ref{sec:system_model} provides a comprehensive overview of the problem formulation. Section \ref{sec:proposed_scheme} explores the underlying algorithmic framework of the proposed DHRL model. Section \ref{sec:sim} presents experimental results. The final section concludes the letter.}

\section{System Model and Problem Formulation}
\label{sec:system_model}

\subsection{Scenario}
\label{subsec:scenario}

\begin{figure}[t]
    \centering
    \includegraphics[draft=false, width= \if 1\mycmd 0.5 \else 0.7 \fi \columnwidth]{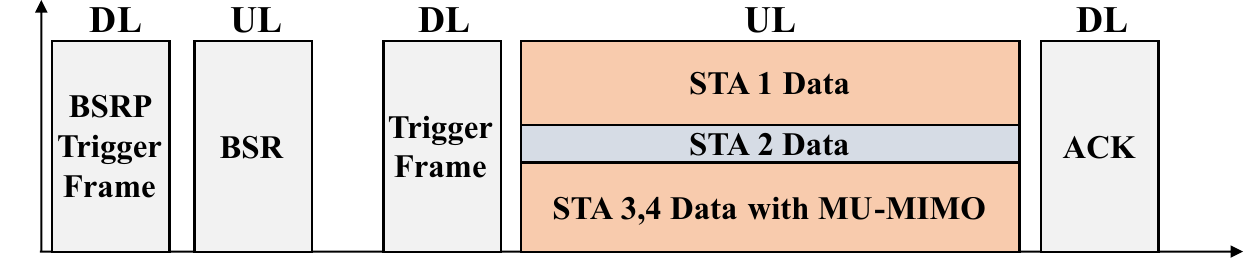}
    \vspace{-7pt}
    \caption{{Illustration of 802.11ax UL data transmission protocol.}}
    \label{fig:protocol}
    \vspace{-7pt}
\end{figure}

In the context of IEEE 802.11ax uplink, we consider a basic service set where an access point (AP) communicates with $K$ stations (STAs). The AP is equipped with $N_\text{R}$ antennas, while each STA employs $N_\text{T}$ antennas. {Fig. \ref{fig:protocol} depicts an illustration of 802.11ax UL data transmission protocol.} The AP estimates the uplink channel state information (CSI) through the uplink pilot. Additionally, the AP gathers buffer status reports (BSR) from the STAs by initiating a BSR poll trigger frame and receiving feedback. With the acquired CSI and buffer status information, the AP determines a combination of resource units (RUs), specifically referring to resource blocks in WLAN, for RA. In addition, the AP schedules single or multiple STAs to transmit uplink data on each RU and specifies the number of transmission packets for individual STAs. Subsequently, the AP acknowledges the scheduled STA by transmitting a trigger frame.

% Then, the AP transmits a trigger frame containing the specifications about the selected STAs' association identifier (AID), the result of RA, and the selected modulation and coding scheme (MCS) for the STAs. After receiving the trigger frame, the STAs check whether they are scheduled or not by matching their own AIDs to the AID12 subfield in the trigger frame. 
% If STAs are scheduled, they encode and transmit uplink data in accordance with the information embedded in the trigger frame through the allocated RUs.

\subsection{OFDMA RU Allocation}
\label{subsec:ofdma_ru_allocation}
\begin{figure}[t]
    \centering
    \includegraphics[draft=false, width= \if 1\mycmd 0.5 \else 0.7 \fi \columnwidth]{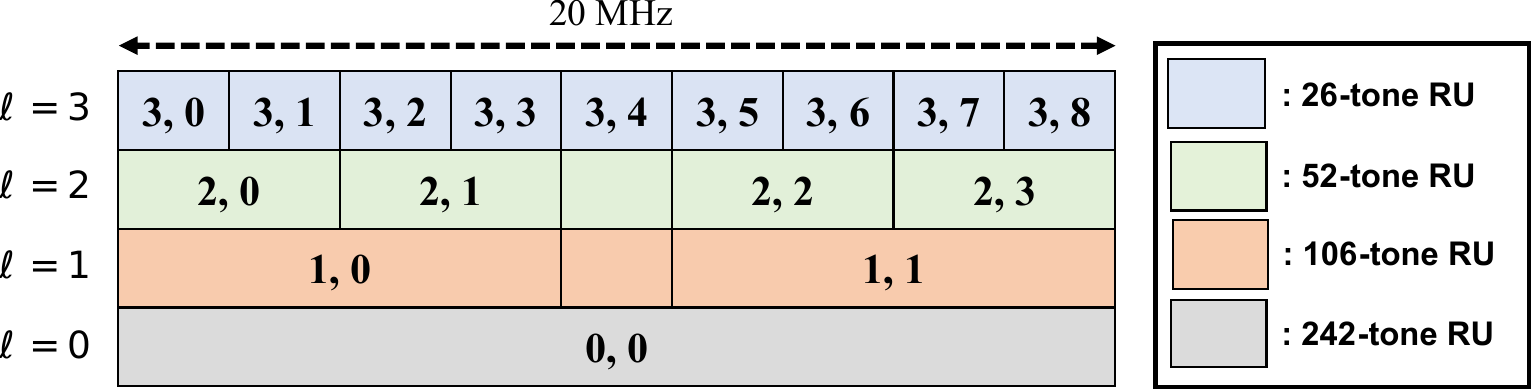}
    \vspace{-7pt}
    \caption{OFDMA RUs in a 20 MHz channel.}
    \label{fig:RU_table}
    \vspace{-12pt}
\end{figure}

According to the IEEE 802.11ax WLAN standard \cite{IEEEstd21}, 256 subcarriers are included in a 20 MHz channel. These 256 subcarriers can be grouped into RUs with various sizes, as depicted in Fig. \ref{fig:RU_table}. 
Let $\text{RU}(l,i)$ denote the RU at the $l$-th level and $i$-th index. Then, each RU can be allocated to an STA for SU-MIMO or multiple STAs for MU-MIMO uplink data transmission. RUs with 106-tone or larger RUs support both SU-MIMO and MU-MIMO, while RUs with 26- or 52-tone permit only SU-MIMO.

The IEEE 802.11ax WLAN standard \cite{IEEEstd21} prohibits an STA from occupying multiple RUs.
Furthermore, the simultaneous allocation of multiple RUs on the same frequency band is forbidden. For example, a valid combination could be $[\text{RU}(1,0),$ $\hspace{-1pt}\text{RU}(3,4),$ $\hspace{-1pt}\text{RU}(2,2),$ $\text{RU}(2,3)]$. To represent the restrictions, we define $\mathbf{e}_{l,i,t} \in \mathbb{Z}^{q \times 1}$ as an indicating vector, the $n$-th element of which is defined such that $[\mathbf{e}_{l,i,t}]_n \hspace{-2pt}=\hspace{-2pt}1$ if $\text{RU}(l,i)$ overlaps with $\text{RU}(L,n)$ in the spectrum domain at the time step $t$, and $[\mathbf{e}_{l,i,t}]_n \hspace{-2pt}=\hspace{-1pt} 0$ otherwise, where {$q$ is the number of 26-tone RUs}, {$[\mathbf{x}]_n$ is the $n$-th entry of a vector $\mathbf{x}$ }, and $L$ is the highest level of RUs (e.g., $L=3$ in the case of a $20~\text{MHz}$ channel).
In addition, $x_{l,i,t}^{(k)}$ is defined as a binary RU assignment variable such that $x_{l,i,t}^{(k)} = 1$ if the $k$-th STA is allocated to $\text{RU}(l,i)$ at the time step $t$, and $x_{l,i,t}^{(k)} = 0$ otherwise.
Then, we can represent the constraints on the RA as follows:
\setlength{\abovedisplayskip}{3pt}
\setlength{\belowdisplayskip}{3pt}
\begin{subequations}\label{eq:constraints}
    \begin{align}
            &\sum_{\forall(l,i)} \delta\left( \sum_{k \in \mathcal{K}} x_{l,i,t}^{(k)} \mathbf{e}_{l,i,t} \right) \preceq \mathbf{1}_p, \label{subeq:1a} \\
            & \sum_{\forall (l,i)} x_{l,i,t}^{(k)} \leq 1, ~~~\forall k \in \mathcal{K}, \label{subeq:1b} \\
            & \sum_{k \in \mathcal{K}} x_{l,i,t}^{(k)} \leq G(l), ~~~\forall (l, i), \label{subeq:1c}
    \end{align}
\end{subequations}
where {$\delta: \mathbb{Z}^{q \times 1} \rightarrow \mathbb{Z}^{q \times 1}$ }is the function defined such that $[\delta(\mathbf{x})]_n = \min\left\{ 
1, \left[\mathbf{x}\right]_n \right\}$ for a vector $\mathbf{x}$, $\mathbf{1}_p$ is the $p$-dimensional vector defined by $\mathbf{1}_p=[1, 1, \ldots, 1]^\text{T}$, $p$ is the number of 26-tone subcarriers, $G$ is the function such that $G(l) = 1$ if $\text{RU}(l,i)$ is either 26- or 52-tone RU, and $G(l) = \lfloor N_\text{R} / N_\text{T} \rfloor$ otherwise, and $\preceq$ is the element-wise inequality operator. Constraint \eqref{subeq:1a} states that two or more different RUs cannot be allocated on the same frequency band. Constraint \eqref{subeq:1b} indicates the unique user assignment along all RUs. Constraint \eqref{subeq:1c} imposes the restriction on MIMO MS, by which MU-MIMO can be employed in RUs with sizes greater than 52-tone.

\subsection{Throughput Model}
\label{subsec:throughput_model}
% , and thereby requires to consider not only STAs' SINRs but also buffer status $\mathbf{b}_t$ as the utility function to focus on more practical network scenarios, where $b_t^{(k)}$ is the number of buffered packets for the $k$-th STA at the time step $t$.
% 우리는 단위 시간당 얼마의 패킷이 실제로 전송되었는지를 측정하기 위하여 throughput을 정의하고 utility function으로 사용한다. 이론적인 utility bound와 다르게, throughput은 STA이 실제로 보내는 패킷의 개수 

Let $\mathbf{H}_{s,t}^{(k)} \in \mathbb{C}^{N_\text{R} \times N_\text{T}}$ denote the uplink communication channel between the AP and the $k$-th STA on the $s$-th subcarrier at the time step $t$. 
After the STAs transmit uplink signals, the AP applies the receive beamformer $\mathbf{w}_{s,t}^{(k)} \in \mathbb{C}^{N_\text{R} \times 1}$ to receive the signal from the $k$-th STA on the subcarrier $s$ at the time step $t$.
Then, the signal-to-interference-plus-noise ratio (SINR) for the $k$-th STA on the subcarrier $s$ at the time step $t$ yields $\Gamma_{s,t}^{(k)} = \frac{P_{s,t}^{(k)} \left\| \left( \mathbf{w}_{s,t}^{(k)} \right)^\text{H} \mathbf{H}_{s,t}^{(k)} \right\|^2_2 }{ \sum_{{m \in \mathcal{K} \backslash \{k\}}} P_{s,t}^{(m)} \left\| \left( \mathbf{w}_{s,t}^{(k)} \right)^\text{H} \mathbf{H}_{s,t}^{(m)} \right\|^2_2  + \sigma^2}$,
% \setlength{\abovedisplayskip}{3pt}
% \setlength{\belowdisplayskip}{1pt}
% \begin{align}
%     \Gamma_{s,t}^{(k)} = \frac{P_{s,t}^{(k)} \left\| \left( \mathbf{w}_{s,t}^{(k)} \right)^\text{H} \mathbf{H}_{s,t}^{(k)} \right\|^2_2 }{ \sum\limits_{\substack{{m \in \mathcal{K} } \\ {m \neq k}}} P_{s,t}^{(m)} \left\| \left( \mathbf{w}_{s,t}^{(k)} \right)^\text{H} \mathbf{H}_{s,t}^{(m)} \right\|^2_2  + \sigma^2},
% \end{align}
where $P_{s,t}^{(k)}$ is the transmit power of the $k$-th STA on the subcarrier $s$ at the time step $t$, $\sigma^2$ is the noise variance, and $\mathcal{K} = \{0,1, \ldots, K-1\}$. {In this letter}, we assume the use of zero-forcing for the receive beamformer. Additionally, we assume $N_\text{T} = 1$ with $\mathbf{h}_{s,t}^{(k)} \in \mathbb{C}^{N_\text{R} \times 1}$ for ease of explanation. 
% Note that since our interest in this letter is to solve the \gls{USRA}, there is no constraint on the beamformer structure. 

We determine the modulation and coding schemes (MCS) of all the STAs according to the IEEE 802.11ax standard \cite{IEEEstd21}. {In particular, we define $m(\Gamma)$ and $c(\Gamma)$ as the modulated bits per symbol and the channel coding rate corresponding to the SINR, respectively. }
Here, it is assumed that channel coherence time is much longer than the OFDM symbol length $\tau$. 
Then, the OFDM rate for the $k$-th STA on $\text{RU}(l,i)$ at the time step $t$ can be given by
\setlength{\abovedisplayskip}{1pt}
\setlength{\belowdisplayskip}{2pt}
% \begin{align}\label{eq:OFDM_rate}
    $O_{l,i}^{(k)} \left(\boldsymbol{\Gamma}_{t} \right) = \sum_{s \in \text{RU}(l,i)} m(\Gamma_{s,t}^{(k)}) c(\Gamma_{s,t}^{(k)}) / \tau,$
% \end{align}
where $\boldsymbol{\Gamma}_{t}$ is the matrix such that $\left[\boldsymbol{\Gamma}_{t}\right]_{s,k} = \Gamma_{s,t}^{(k)}$. 
With the determined OFDM rate, the $k$-th STA transmits $p_t^{(k)}$ packets within the limited physical protocol data unit (PPDU) time $\tau_\text{PPDU}$. Clearly, the transmission time for $p_t^{(k)}$ packets can be represented by $T_{l,i}^{(k)}(\boldsymbol{\Gamma}_t, p_t^{(k)}) = Q p_t^{(k)} / O_{l,i}^{(k)}(\boldsymbol{\Gamma}_t)$, where $Q$ is the length of each packet. The data throughput (bits/s) of the $k$-th STA on $\text{RU}(l,i)$ \cite{Ha21} can be represented by 
\begin{align}
\label{throughput_for_STA}
    R_{l,i}^{(k)} \hspace{-1pt} (\mathbf{X}_t, \boldsymbol{\Gamma}_t, \mathbf{p}_t) \hspace{-1pt}=\hspace{-1pt} \frac{ x_{l,i,t}^{(k)} O_{l,i}^{(k)} \left( \boldsymbol{\Gamma}_{t} \right) T_{l,i}^{(k)}\left( \boldsymbol{\Gamma}_{t}, p_{t} \right) }{\max_{l,i,k} x_{l,i,t}^{(k)} \hspace{-1pt} \left( T_{l,i}^{(k)} \hspace{-2pt} \left( \boldsymbol{\Gamma}_{t}, p_{t} \right) \hspace{-1pt} + \hspace{-1pt} V^{(k)} \hspace{-1pt} \right)},
\end{align} 
where $\mathbf{X}_t$ is the three-dimensional data cube such that $\left[\mathbf{X}_t \right]_{l,i,k} = x_{l,i,t}^{(k)}$, $\mathbf{p}_t = [p_t^{(0)}, \ldots, p_t^{(K-1)}]$, and $V^{(k)}$ is the overhead for the $k$-th STA.

\subsection{Joint \gls{USRA} Optimization Problem}
Considering the constraints on the RA and packet transmission, we formulate the \gls{USRA} optimization problem as follows: 
\setlength{\abovedisplayskip}{-5pt}
\setlength{\belowdisplayskip}{2pt}
\begin{subequations}
\label{P:opt}
\begin{align}
    & \max_{\mathbf{X}_t, \mathbf{p}_t} & & \sum_t \sum_k \sum_{\forall (l,i)} R_{l,i}^{(k)}({\mathbf{X}}_{t}, \boldsymbol{\Gamma}_{t}, \mathbf{p}_{t}), \\
    & \text{s.t.} & & \eqref{subeq:1a} \text{-} \eqref{subeq:1c}, \\
    & & & p_t^{(k)} < b_t^{(k)}, \forall k \in \mathcal{K}, \label{subeq:buffer} \\
    & & & \frac{Qp_t^{(k)}}{O_{l,i}^{(k)}(\boldsymbol{\Gamma}_t)} < \tau_\text{PPDU}, \forall (l,i), \forall k \in \mathcal{K}, \label{subeq:PPDU}
\end{align}
\end{subequations}
where $b_t^{(k)}$ is the buffer status of the $k$-th STA at the time step $t$.
The aim of the problem \eqref{P:opt} is to maximize overall data throughput while adhering to the constraint on RA and MIMO MS in \eqref{eq:constraints}. Furthermore, the number of transmission packets should be judiciously determined to ensure that STAs do not transmit packets exceeding their buffer status in \eqref{subeq:buffer}, and their transmission does not surpass the maximum PPDU time in \eqref{subeq:PPDU}.
Solving the \gls{USRA} optimization problem (\ref{P:opt}) proves to be intractable, given its nature as mixed-integer non-linear programming, which has been established as NP-hard. Notably, the \gls{USRA} optimization problem (\ref{P:opt}) incorporates aspects such as OFDMA, MU-MIMO, MIMO MS, and buffer status of the STAs, rendering it more challenging compared to previous works that did not account for all these facets.

\section{Proposed DHRL Algorithm}
\label{sec:proposed_scheme}

\begin{figure}[t]
    \centering
    \includegraphics[draft=false, width= \if 1\mycmd 0.7 \else 0.8 \fi \columnwidth]{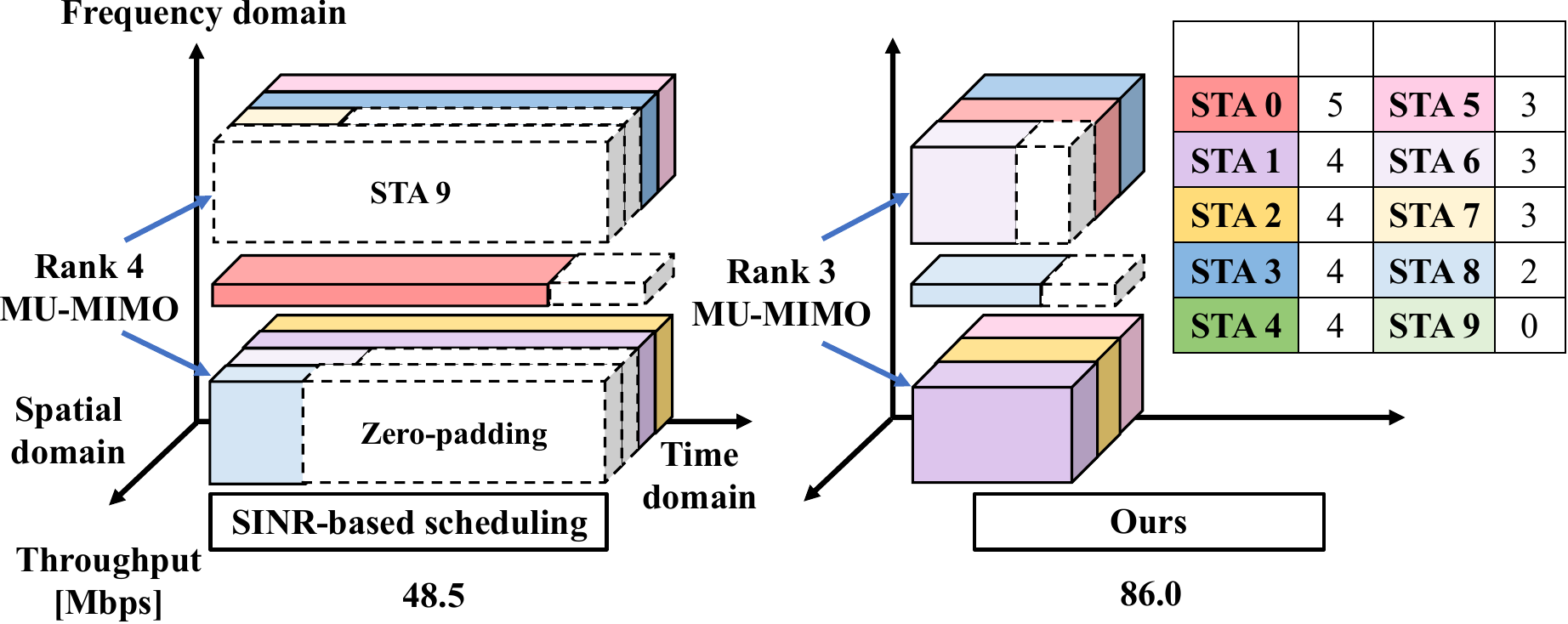}
    \vspace{-10pt}
    \caption{An illustrative example of the \gls{USRA} when scheduling STAs and allocating RUs with conventional and proposed methods.}
    \label{fig:buffer_ex}
    \vspace{-7pt}
\end{figure}

\subsection{Overview of The Proposed Scheme}
Fig. \ref{fig:buffer_ex} illustrates the importance of considering the buffer status of each STA under unsaturated traffic conditions. 
The left figure shows inefficiencies of utilizing the semiorthogonal user selection (SUS) algorithm \cite{Yoo06} for MU-MIMO user selection and state-of-the-art uplink RA techniques \cite{Wang18} in such conditions.
Both algorithms, though near-optimal in saturated traffic conditions by focusing on SINR, lead to inefficiencies when STAs with minimal or empty buffer status but good SINR are scheduled alongside other STAs with large buffer status on the same RU. Hence, long periods of redundant zero padding may be inserted to wait until the end of the other STAs' data transmission. Furthermore, it favorably assigns high-rank MU-MIMO to maximize the sum of SINR. 
The right figure serves as an example of solving the \gls{USRA} problem using the proposed DHRL-based method. 
This method optimizes MU-MIMO user selection and RA along the time-frequency-space axis by factoring in both the STAs' SINRs and buffer status, so that it allows for efficient RA and MU-MIMO designs sometimes with low-rank MU-MIMO, instead of always relying on high-rank MU-MIMO.

\subsection{Structure and parameters of the proposed DHRL}
\label{subsec:proposed_DHRL_structure}
% \begin{figure}[t]
%     \centering
%     \includegraphics[draft=false, width= \if 1\mycmd 0.7 \else 0.8 \fi \columnwidth]{Figures/HDQN.pdf}
%     \vspace{-8pt}
%     \caption{The structure of the proposed DHRL for a 20 MHz channel.}
%     \label{fig:HDQN}
%     \vspace{-15pt}
% \end{figure}

\begin{figure}[t]
\centering
\subfigure[]{
\includegraphics[draft=false, width= \if 1\mycmd 0.4 \else 0.56 \fi \columnwidth]{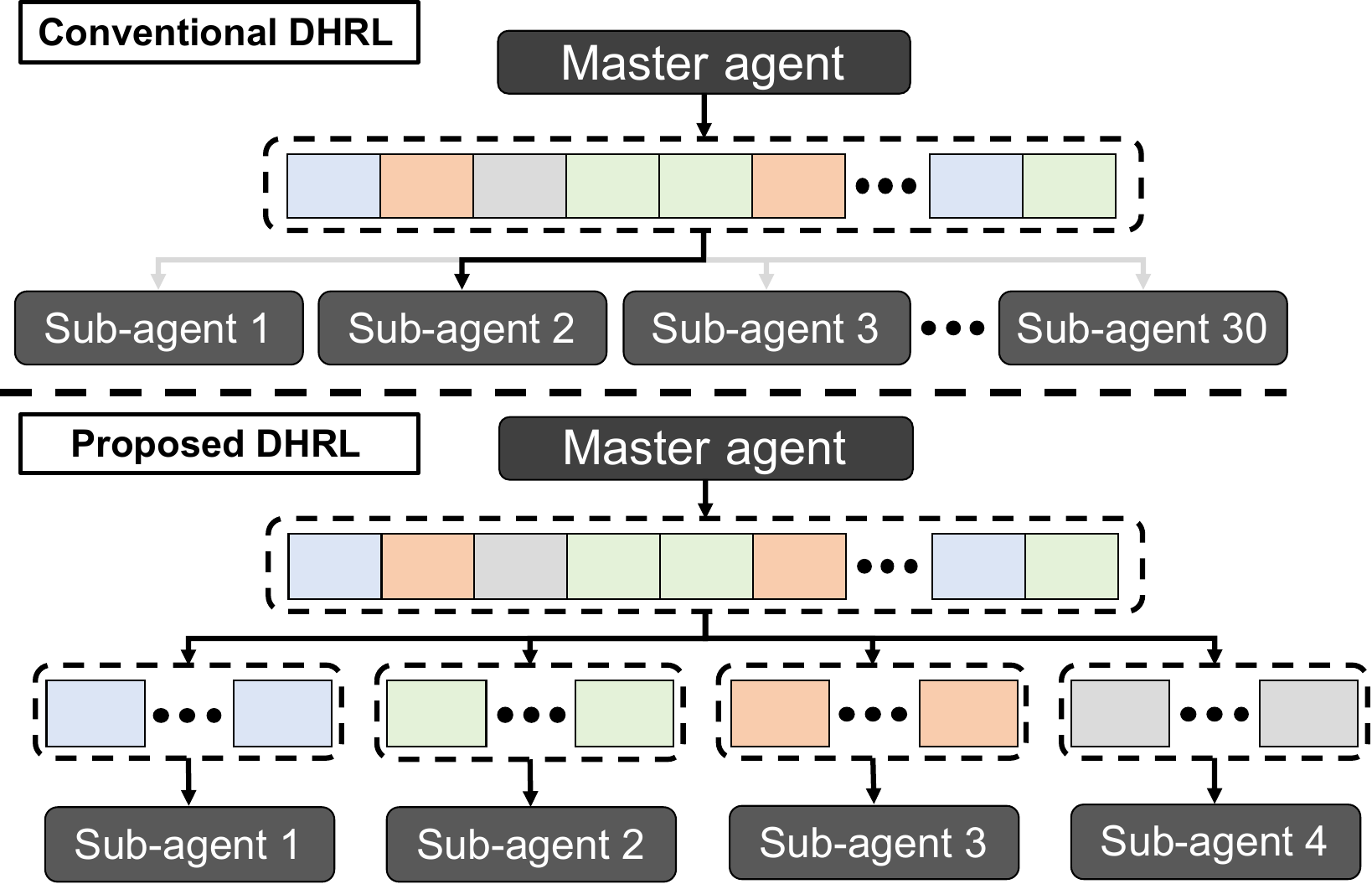}
\label{fig:conv_proposed}
}
\subfigure[]{
\includegraphics[draft=false, width= \if 1\mycmd 0.4 \else 0.36 \fi \columnwidth]{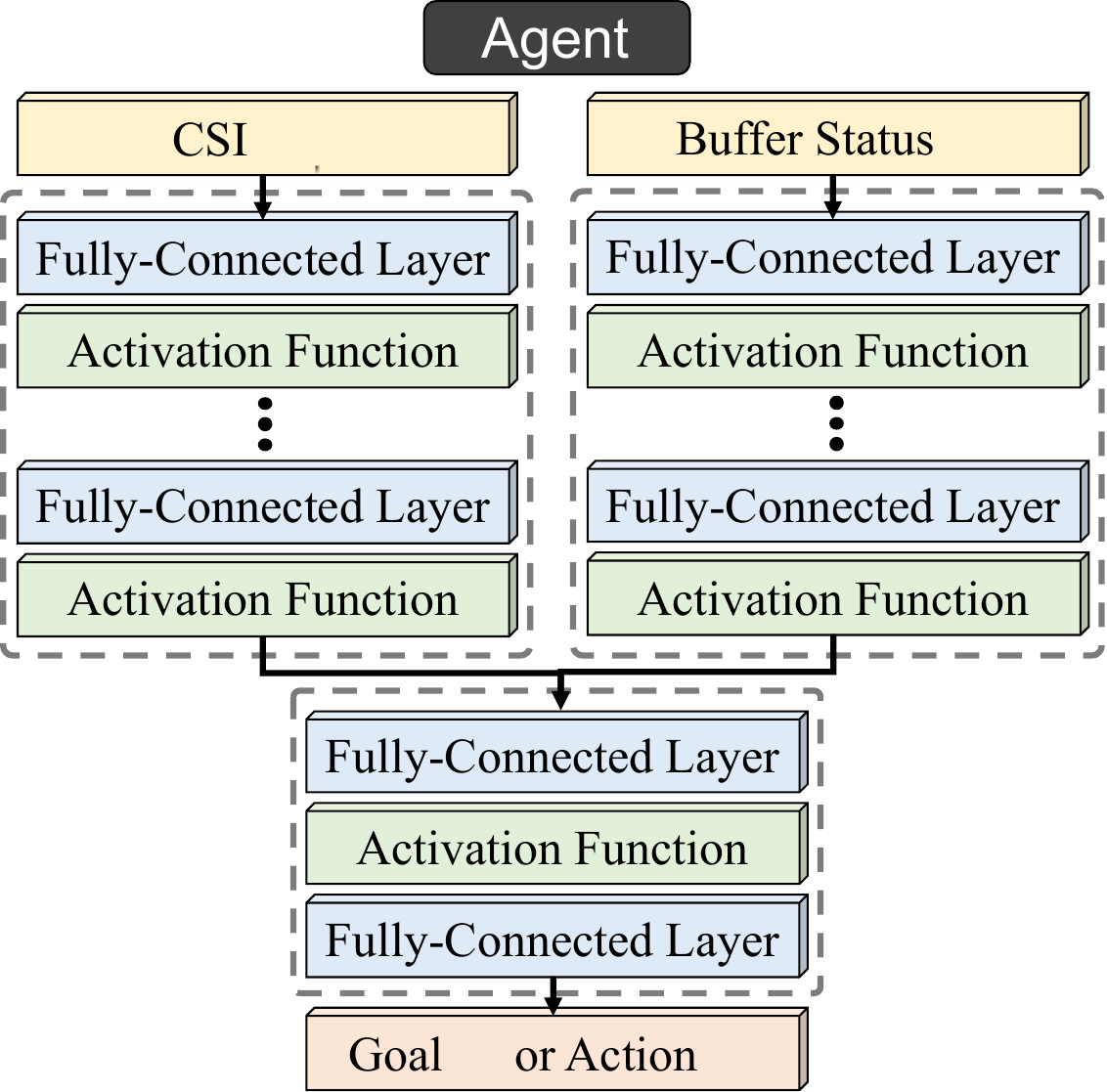}
\label{fig:network}
}
\vspace{-6pt}
\caption{Illustration of the proposed DHRL model structure. (a) The comparison between the conventional and proposed DHRL model in a 20 MHz channel. {A master agent performs RA, and sequentially sub-agents schedule STAs on the allocated RUs. Then, the AP transmits a trigger frame containing the result of \gls{USRA} to the STAs. }(b) The network structure of both master and sub-agents.}
\label{fig:proposed_DHRL}
\vspace{-6pt}
\end{figure}

{Decomposing the \gls{USRA} problem into a hierarchy of RA and US, an AP defines master and sub-agents on its own, without collaboration with other nodes, each designed for solving RA and US problems, respectively. Here, we define the result of RA as a goal and the result of US as an action. The \gls{USRA} process begins with the master agent setting a goal $g_t$. Unlike conventional DHRL approaches that assign all RUs to a single sub-agent, our model is tailored to the frequency band use cases in 802.11ax. This involves decomposing the RU combination determined by $g_t$ into individual units based on size, as shown in Fig. \ref{fig:conv_proposed}. These segmented RUs are inputs for sub-agents, who select actions $a_t$ based on the current state and goal. This refined sub-agent structure profoundly optimizes network\footnote{{For example, the sub-agent design reduces the number of sub-agents from 30 to 4 and their action space sizes from $9.2 \times 10^{10}$ to $6.2 \times 10^{3}$ for scenarios with four antennas and 20 STAs in a 20 MHz channel.}}, enhancing network performance and convergence speed. 
Finally, the AP constructs $\mathbf{X}_t$ with the actions and transmits it to the STAs in a trigger frame.}

% To begin our DHRL model training, the master agent first determines a goal $g_t$. 
% Rather than directly allocating the whole combination of RUs to a single sub-agent like the conventional DHRL, we tailor the DHRL model to the MU-MIMO user selection, as shown in Fig. \ref{fig:conv_proposed}. Hence, the combination of RUs decided by $g_t$ is decomposed into RUs according to their sizes. The classified RUs are used as the inputs for the sub-agents. Then, the sub-agents select actions $a_t$ based on the current states and goal. This novel sub-agent design profoundly optimizes the network\footnote{{For example, The sub-agent design reduces the number of sub-agents from 30 to 4 and their action space from $9.2 \times 10^{10}$ to $6.2 \times 10^{3}$ for scenarios with four antennas and 20 STAs in a 20 MHz channel. }}, enhancing network performance and convergence speed.}
% Finally, after taking the actions, the next states and rewards are fed back into the master agent and sub-agents.

{Neural networks typically expect their input data to be homogeneous in terms of format, structure, and characteristics for a given task. However, we hope a single agent to concurrently process both CSI and buffer status, which exhibit heterogeneous characteristics. To this end, we design an agent with two branch neural networks, as shown in Fig. \ref{fig:network}, each tailored to process either CSI or buffer status, before merging them into the fusion network. This architecture allows for the effective integration of diverse data types, enabling the agent to make more informed decisions based on a comprehensive understanding of the network state.}

\begin{figure}[t]
    \centering
    \includegraphics[draft=false, width= \if 1\mycmd 0.7 \else 0.8 \fi \columnwidth]{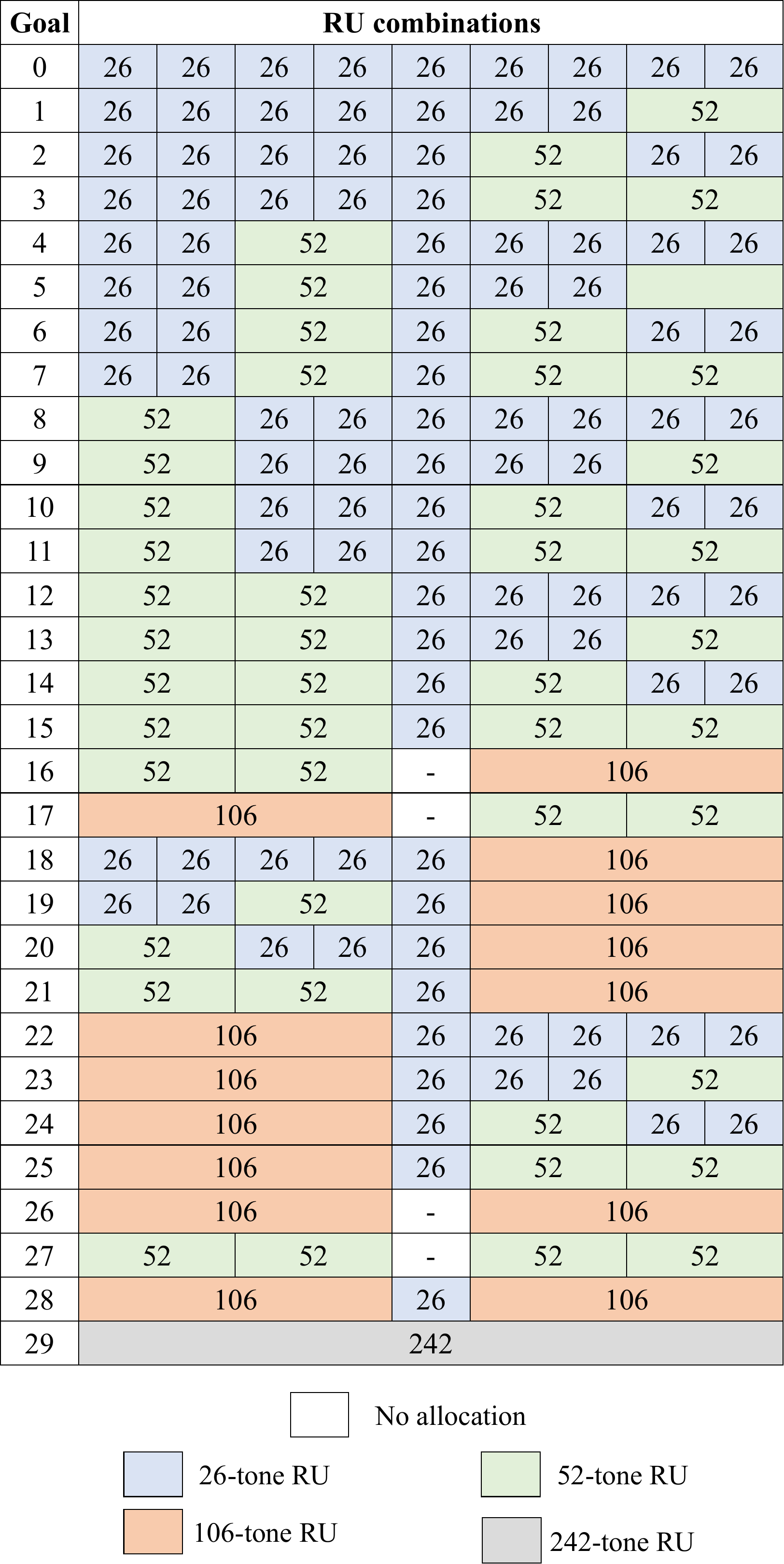}
    \caption{RU combination table corresponding to goals in a 20 MHz bandwidth channel.}
    \label{fig:goal_table}
\end{figure}

The state, goal, action, and reward functions of the master agent and sub-agents are defined below.
\begin{itemize}
    \item \textit{State}: The state for the $l$-th level and $i$-th index at the time step $t$ is $s_t^{(l,i)} \hspace{-5pt} = \hspace{-3pt} \{  \mathbf{H}_{s,t}^{(k)}, b_{t}^{(k)} | \forall s\in \text{RU}(l,i), \forall k \in \mathcal{K} \}$. Then, the state for all levels and indices at the time step $t$ is $\mathcal{S}_t = $ $ \bigcup_{l=0}^L \bigcup_{i \in \mathcal{I}(l)} s_t^{(l,i)}$, where $\mathcal{I}(l)$ is the set containing the integer from 0 to the maximum index at the $l$-th level.
    \item \textit{Goal}: The outcome determined by a master agent regarding the RA is referred to as the goal, and is defined as $g_t \in \mathcal{G}$, where $\mathcal{G}$ is the goal space including all possible cases of RA according to the IEEE 802.11ax standard \cite{IEEEstd21}. 
    A master agent selects a goal, which determines the resource allocation format based on the RU combination table in Fig. \ref{fig:RU_table}. For example, if a master agent selects goal 21, the RU combination is $[\text{RU}(2, 0), \text{RU}(2, 1),$ $ \text{RU}(3, 4), \text{RU}(1, 1)]$. Subsequently, sub-agents perform user scheduling on each RU. Therefore, sub-agents perform user scheduling on the RU combination corresponding to the selected goal $g_t$, $[\text{RU}(l_0, i_0), \text{RU}(l_1, i_1), \ldots, \text{RU}(l_d, i_d)]$.
    \item {\textit{Action}: At the time step $t$, we define an action as $a_{t}^{(l,i)} \in \mathcal{A}^{(l,i)} = \{ S | S \subset \mathcal{K}, |S| \leq G(l) \}$, where $\mathcal{A}^{(l,i)}$ is the action space on $\text{RU}(l,i)$. The action $a_t^{(l,i)}$ specifies the STAs to be scheduled on $\text{RU}(l,i)$ at the time step $t$, i.e, $x_{l,i,t}^{(k)} = 1$ if $k \in a_t^{(l,i)}$. Then, sub-agents supporting MU-MIMO have action spaces of size $\sum_{i=1}^{\floor{N_\text{R} / N_\text{T}}}$ $ {K}\choose{i}$ (constraint \eqref{subeq:1c}) respectively which are still heavy compared to SU-MIMO. To further reduce the sub-agents' action space sizes, we sophisticatedly redesign the action of sub-agents in section \ref{subagent_design}.}
    \item \textit{Reward Function}: After all sub-agents finish the action selection process, a master agent constructs $\mathbf{X}_t$ with $a_t^{(l,i)}$ for all levels and indices, executes $\mathbf{X}_t$, and obtains extrinsic reward, which is defined by $\hat{r}_t = $ $\sum_{l=0}^L \sum_{i\in \mathcal{I}(l)} \sum_{k\in\mathcal{K}} R_{l,i}^{(k)} (\mathbf{X}_{t}, \boldsymbol{\Gamma}_t, \mathbf{p}_t)$. Given $\mathbf{X}_t$, $\mathbf{p}_t$ can be constituted using a simple search algorithm with a computational complexity of $\mathcal{O}(\left\| \mathbf{X}_t \right\|_\text{F})$, where $\left\| \mathbf{X}_t \right\|_\text{F}$ represents the number of scheduled STAs. {Reward function for sub-agents is defined in section \ref{subagent_design}.}
\end{itemize}

\subsection{Sub-agent design for MU-MIMO}
\label{subagent_design}

% To reduce the action space sizes of sub-agents to a level comparable to SU-MIMO, we define the decision round $\tilde{t}$ within a time step $t$ and design sub-agents that repeatedly select one STA at each decision round for at most $\floor{N_\text{R} / N_\text{T}}$ times.
% Additionally, at each decision round, the agent can take a "break" option, which allows for the early stop of the action selection procedure to select fewer actions than $\floor{N_\text{R} / N_\text{T}}$.
% Then, we significantly reduce the action space sizes from $\sum_{i=1}^{\floor{N_\text{R} / N_\text{T}}}$ $ {K}\choose{i}$ to $K+1$, which boosts the convergence speed of the sub-agents and improves network performance.

% 기존에 정의된 huge action space 사이즈는 MU-MIMO user selection의 모든 경우의 수를 한번에 고려하는 것을 원인으로 한다. 우리는 대신에 one time step 내에 최대 $\floor{N_\text{R} / N_\text{T}}$ 만큼의 decision round를 정의하고, 각 decision round마다 하나의 STA만을 선택하게 한다. 또한 $\floor{N_\text{R} / N_\text{T}}$보다 적은 유저 선택을 고려하기 위하여 각 decision round마다 "break" option을 선택할 수 있으며, which allows for the early stop of the decision round. 

The huge size of previously defined action space stems from the practice of contemplating every possible permutation for MU-MIMO user selection at once. Instead, we propose a sequential user selection approach where $\floor{N_\text{R} / N_\text{T}}$ decision rounds are sequentially conducted within a single time step, during which sub-agents select one STA per round. To account for scenarios where selecting fewer users than $\floor{N_\text{R} / N_\text{T}}$ is preferable, sub-agents can choose a `break' option in any decision round, allowing for the early termination of the selection process.
Via the proposed user selection approach, we significantly reduce the action space sizes from $\sum_{i=1}^{\floor{N_\text{R} / N_\text{T}}}$ $ {K}\choose{i}$ to $K+1$, which boosts the convergence speed of sub-agent training and improves network performance.

We redefine an action as $\tilde{a}_{\tilde{t}}^{(l,i)} \in \mathcal{K} \cup \{break\} = \tilde{\mathcal{A}}$, where $break$ represents the break option, and $\tilde{\mathcal{A}}$ is the newly defined action space. Across multiple decision rounds, each sub-agent iteratively selects actions. Subsequently, the selected actions are aggregated to form $a_t^{(l,i)} = \{ \tilde{a}_0^{(l,i)}, \ldots, \tilde{a}_{T-1}^{(l,i)} \}$, where $T$ is the terminal time step.
After decision rounds finish, each sub-agent constitutes $\mathbf{X}_t$ with $a_t^{(l,i)}$ executes $\mathbf{X}_t$, and obtains intrinsic reward, $r_t^{(l,i)} = \sum_{k\in\mathcal{K}} R_{l,i}^{(k)} (\mathbf{X}_{t}, \boldsymbol{\Gamma}_t, \mathbf{p}_t)$. 

In contrast to SU-MIMO, MU-MIMO can lead to reduced throughput if there is strong mutual interference among STAs. Therefore, we design a model structure aimed at mitigating mutual interference among selected STAs. 
To this end, we continuously update and provide the orthogonal components of the STAs' channel with respect to the selected STAs' channels at each decision round, incorporating this information into the sub-agents' states.
% To this end, we continuously update and provide the orthogonal components of the STAs' channel with respect to the selected STAs' channels at each decision round, incorporating this information into the sub-agents' states\footnote{With zero-forcing receiver, mutual interference among STAs can be reduced by selecting a group of STAs with high channel magnitudes, whose channel directions align with the zero-forcing beam directions \cite{Yoo06}.}.
Thus, we repeatedly calculate $\mathbf{g}_{s,t, \tilde{t}}^{(k)}$, the component of the STAs' channels orthogonal to the subspace spanned by $\left\{ \tilde{\mathbf{g}}_{s,t,1}, \ldots, \tilde{\mathbf{g}}_{s,t,\tilde{t}-1} \right\}$ at each decision round $\tilde{t}$, which can be obtained by
\setlength{\abovedisplayskip}{3pt}
\setlength{\belowdisplayskip}{1pt}
\begin{align}\label{eq:span_update}
    \mathbf{g}_{s,t,\tilde{t}}^{(k)} = \mathbf{h}_{s,t}^{(k)} - \sum_{j=1}^{\tilde{t}-1} \frac{\left(\mathbf{h}_{s,t}^{(k)}\right)^\text{H} \tilde{\mathbf{g}}_{s,t,j}}{\| \tilde{\mathbf{g}}_{s,t,j} \|^2} \tilde{\mathbf{g}}_{s,t,j},
\end{align}
where $\mathbf{g}_{s,t,1}^{(k)} = \mathbf{h}_{s,t}^{(k)}$ for all $k \in \mathcal{K}$, $\tilde{\mathbf{g}}_{s,t,j} = \mathbf{g}_{s,t,j}^{(k^{'})}$, and $k^{'}$ is the index of selected STA at the $j$-th decision round. Then, we utilize $\mathbf{g}_{s,t,\tilde{t}}^{(k)}$ as new states for MU-MIMO. 
\begin{algorithm}[t]
\caption{Proposed DHQN-based \gls{USRA} algorithm}
\label{alg:HDQN_USRA}
\begin{algorithmic}[1]
\State \textbf{Initialize: } $\theta_{\text{M}}$, $\theta_{\text{S}}^{(l)}$, $\mathcal{D}_\text{M}$,
    $\mathcal{D}_{\text{S}}^{(l)}$,
    $s_t^{(l,i)}, \forall (l,i).$
    
\For {$e = 1, ...,num\_episodes$}
    \State Get start state description $\mathcal{S}_0$, $t \leftarrow 0$

    \While {$\mathcal{S}_t$ is \textbf{not} terminal}
        % \State $g_t \leftarrow $ \Call{EpsilonGreedy}{$\mathcal{S}_t, \mathcal{G}, e, \theta_\text{M}$}
        \State {Select $g_t$ with a $\epsilon$-greedy policy}
        \For{$j=0,...,d$}
            % \State $a_t^{(l_j, i_j)} \leftarrow $ EpsilonGreedy$(\{s_t^{(l_j,i_j)}, g_t \}, \mathcal{A}^{(l_j)}, e, \theta_\text{S}^{(l_j)})$
            \State $ a_t^{(l_j, i_j)} \leftarrow $ \Call{ActionSelection}{$\mathcal{S}_t, (l_j, i_j), e$}
        \EndFor
        \State Execute $a_t$, and obtain $\hat{r}_t$ and $\mathcal{S}_{t+1}$
        \State Store $\left( \mathcal{S}_t, g_t, \hat{r}_t, \mathcal{S}_{t+1} \right)$ in $\mathcal{D}_\text{M}$
        \State {Update $\theta_\text{M}$ by the stochastic gradient descent method}
        % \State Execute $a_t$, and obtain $\hat{r}_t$ and $\mathcal{S}_{t+1}$
        % \State Store $\left( \mathcal{S}_t, g_t, \hat{r}_t, \mathcal{S}_{t+1} \right)$ in $\mathcal{D}_\text{M}$
        % \State {Update $\theta_\text{M}$ by the stochastic gradient descent method}
        % \State Sample mini-batch of $B$ transitions from $\mathcal{D}_\text{M}$
        % \State Update $\theta_{\text{M}}$ to minimize $J(\theta_{\text{M}})$
        \State $t \leftarrow t+1$
    \EndWhile
\EndFor
\State \textbf{return} $\mathbf{X}_t$
\end{algorithmic}
\end{algorithm}
\setlength{\textfloatsep}{5pt}
\subsection{DHRL Training Process}
\label{subsec:DHRL_training_propcess}
We introduce the Deep Q-network (DQN) as the learning framework of the master and sub-agents. Algorithm \ref{alg:HDQN_USRA} describes the proposed deep hierarchical Q-network (DHQN)-based \gls{USRA}. With the newly-defined action $\tilde{a}$ and action space $\tilde{\mathcal{A}}$ in section \ref{subagent_design}, we train a DHRL model as follows.
\subsubsection{Initialization}
We utilize a parameter $\theta_\text{M}$ that defines an action-value function $Q(\mathcal{S},g ; \theta_\text{M})$ for the master agent and parameters $\theta_{\text{S}}^{(l)}$ that define action-value functions $Q(s,a;\theta_\text{S}^{(l)}, g)$ for the $l$-th level sub-agent, where $l$ ranges from $0$ to $L$. In addition, we initialize replay memories $\mathcal{D}_\text{M}$ for the master agent to capacity $C_\text{M}$ and $\mathcal{D}_{\text{S}}^{(l)}$ for the $l$-th sub-agent to capacity $C_{\text{S}}^{(l)}$.

\subsubsection{Experience collection}
\label{subsubsec:experience_collection}
{The master agent first selects a goal $g_t$ with a $\epsilon$-greedy policy, where $\epsilon$ decays from 1.0 to 0.1.} For $[\text{RU}(l_0, i_0)$, $\hspace{-1pt} \text{RU}(l_1, i_1), \ldots,\text{RU}(l_d, i_d)]$ determined by the selected goal $g_t$, each sub-agent selects an action $a_t^{(l_j, i_j)}$.
Algorithm \ref{alg:actselection} shows the action selection process in the sub-agents. 
A sub-agent supporting MU-MIMO repeatedly appends an action $\tilde{a}_{\tilde{t}}^{(l_j,i_j)}$ to $a^{(l_j, i_j)}$ for at most $\floor{N_\text{R} / N_\text{T}}$ decision rounds. At each decision round $\tilde{t}$, the sub-agent evaluates the reward difference $r_{\tilde{t}}^{(l_j, i_j)} - r_{\tilde{t}-1}^{(l_j, i_j)}$ and stores transition $\left( \tilde{s}_{\tilde{t}}^{(l_j,i_j)}, \tilde{a}_{\tilde{t}}^{(l_j,i_j)}, r_{\tilde{t}}^{(l_j,i_j)} - r_{\tilde{t}-1}^{(l_j,i_j)}, \tilde{s}_{\tilde{t}+1}^{(l_j,i_j)} \right)$ in $\mathcal{D}_\text{S}^{(l_j, i_j)}$.
The master agent also stores transition $\left( \mathcal{S}_t, g_t, \hat{r}_t, \mathcal{S}_{t+1} \right)$ in $\mathcal{D}_\text{M}$ at each time-step $t$.

\subsubsection{Updating model parameters}
\label{subsubsec:updating_model_params}
{In each training epoch, we randomly sample mini-batches from the datasets and update the model parameters, $\theta_\text{M}$ and $\theta_\text{S}^{(l)}$, by the stochastic gradient descent method. In the case of $\theta_\text{M}$, we set $y_j = \hat{r}_j$ if $\mathcal{S}_{j+1}$ is a terminal state or $y_j = \hat{r}_j + \gamma \max_{g} Q(\mathcal{S}_{j+1}, g; \theta_\text{M})$, otherwise. 
Then, the loss function for a given mini-batch $\mathcal{B}$ is defined as $J(\theta_{\text{M}}) = \sum_{j\in \mathcal{B}} \left( y_j-Q(\mathcal{S}_j, g_j; \theta_{\text{M}}) \right)^2 / |\mathcal{B}|$. 
By the stochastic gradient descent method, the parameter $\theta_\text{M}$ is updated to minimize the loss function, which yields
\begin{align}
    \theta_\text{M} \leftarrow \theta_\text{M}-\frac{\alpha_\text{M}}{|\mathcal{B}|} \frac{\partial J(\theta_\text{M})}{\partial \theta_\text{M}},
\end{align}
where $\alpha_\text{M}$ is the learning rate. In the same way, the sub-agents can update $\theta_{\text{S}}^{(l)}$.}

\begin{algorithm}[t]
\caption{\textsc{ActionSelection($\mathcal{S}_t, (l, i), e$)}}
\label{alg:actselection}
\begin{algorithmic}[1]
\IIf{$l \leq 1$} $M \leftarrow \floor{N_\text{R}/N_\text{T}}$ \ElseIIf{$M \leftarrow 1$}\EndIIf
\State $\tilde{s}_{0}^{(l,i)} \leftarrow s_{t}^{(l,i)}$, $r_{-1}^{(l,i)} \leftarrow 0$, $a^{(l,i)} \leftarrow [~]$
\For{$\tilde{t} = 0, ..., M-1$}
\State {Select $\tilde{a}_{\tilde{t}}^{(l,i)}$ with a $\epsilon$-greedy policy}
% \State $\tilde{a}_{\tilde{t}}^{(l,i)} \leftarrow$ \Call{EpsilonGreedy}{$\tilde{s}_{\tilde{t}}^{(l,i)}, \mathcal{A}, e, \theta_\text{S}^{(l)}$}  
% \If {$\tilde{a}_{\tilde{t}}^{(l,i)} = break$}
%     \State break
% \EndIf
\IIf{$\tilde{a}_{\tilde{t}}^{(l,i)} = break$} break \EndIIf
% \State $a^{(l,i)}[\tilde{t}] \leftarrow \tilde{a}_{\tilde{t}}^{(l,i)} $
\State Append $\tilde{a}_{\tilde{t}}^{(l,i)} $ to $a^{(l,i)}$
\State Execute $a^{(l,i)}$, obtain $r_{\tilde{t}}^{(l, i)}$, and update $\tilde{s}_{\tilde{t}+1}^{(l,i)}$ from \eqref{eq:span_update}
\State Store $\left( \tilde{s}_{\tilde{t}}^{(l,i)}, \tilde{a}_{\tilde{t}}^{(l,i)}, r_{\tilde{t}}^{(l,i)} - r_{\tilde{t}-1}^{(l,i)}, \tilde{s}_{\tilde{t}+1}^{(l,i)} \right)$ in $\mathcal{D}_{\text{S}}^{(l)}$
\State {Update $\theta_\text{S}^{(l)}$ by the stochastic gradient descent method}
% \State Sample mini-batch of $B$ transitions from $\mathcal{D}_\text{S}^{(l)}$
% \State Update $\theta_{\text{S}}^{(l)}$ to minimize loss 
\EndFor
\State \textbf{return} {$a^{(l,i)}$}
\end{algorithmic}
\end{algorithm}
\setlength{\textfloatsep}{5pt}
{
\subsection{Complexity analysis}
\label{subsec:complexity_analysis}
We find that the integrated part from the branch network for CSI to the fusion network dominates the majority of computational complexity. Thus, we focus on the computational complexity of such a part. Defining the network depth of the integrated part as $d$, with the input size of $K \cdot 2^{L/2}$, the computational complexity is $\mathcal{O}(d K^2 2^L)$. In addition, the state update process in \eqref{eq:span_update} requires $\mathcal{O}(MK)$ computation, where $M$ is the maximum iteration level. Then, the computational complexity of the proposed DHRL algorithm reads $\mathcal{O}(d MK^3 2^L)$. On the other hand, the conventional work \cite{Wang18} requires $\mathcal{O}(MK^2 4^L)$. While our DHRL algorithm requires $d K$ times more computational effort, it benefits exponentially from wider bandwidths, as reflected by the more favorable exponential term $2^L$ compared to $4^L$ in the conventional work.
}
\section{Simulation Results} \label{sec:sim}
We have conducted simulations to evaluate the performance of the proposed DHQN-based \gls{USRA} algorithm and compare it with baseline methods.
{A 802.11ax system, where an AP is equipped with $N_\text{R} = 8$ antennas and STAs employ $N_\text{T} = 2$ antennas, is considered with the bandwidth $B$ set to $20~\text{MHz}$. }
For our simulation, we build a MATLAB simulator based on the IEEE standard channel model document \cite{80211_channel}. The STAs are dropped randomly within the distance range of {$20~\text{m}$ to $100~\text{m}$.}
Packet length $Q$ is $1500~\text{bytes}$, the maximum PPDU time is $4.848~\text{ms}$, and packet arrivals follow the Poisson distribution. 

For comparison, we consider three baseline \gls{USRA} methods. 
\begin{itemize}
    \item SINR-based scheduling: This method optimizes \gls{USRA} by adopting the algorithms in \cite{Wang18} and \cite{Yoo06}, which are proven to be near-optimal solutions under saturated traffic conditions.
    \item SINR-based scheduling (fixed RA): It maximizes the STAs' throughput by optimizing US but with fixed RA. If MU-MIMO and OFDMA are utilized, the algorithm chooses level $l = \min(L-2, \floor{\log_2 (K \cdot N_\text{R} / N_\text{T})}) $ \cite{Wang18}.
    \item Buffer-based scheduling (fixed RA): We design a heuristic algorithm that chooses the same combination of RUs in the SINR-based scheduling (fixed RA) algorithm and assigns as many STAs as possible on every RU. The algorithm selects STAs in ascending order of the number of buffers $b_t^{(k)}$. 
\end{itemize}

\setlength{\textfloatsep}{20pt}
\begin{figure}[t]
\centering
\subfigure[]{
\includegraphics[draft=false, width= \if 1\mycmd 0.36 \else 0.45 \fi \columnwidth]{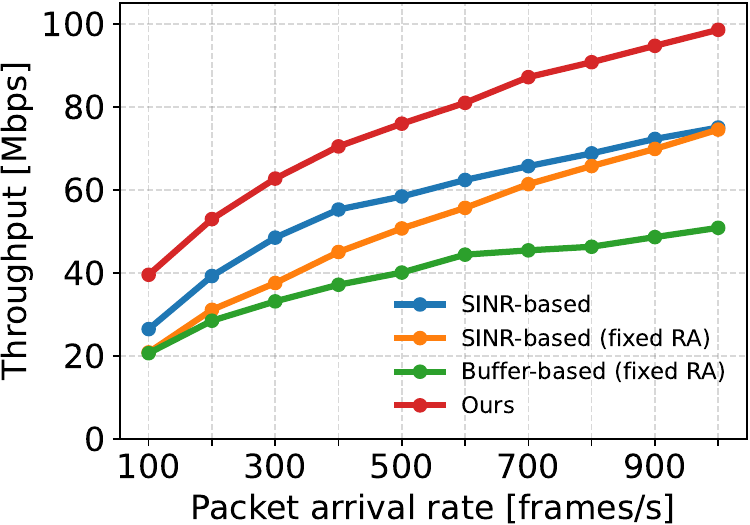}
\label{fig:throughput_packet_rate}
}
\subfigure[]{
\includegraphics[draft=false, width= \if 1\mycmd 0.36 \else 0.45 \fi \columnwidth]{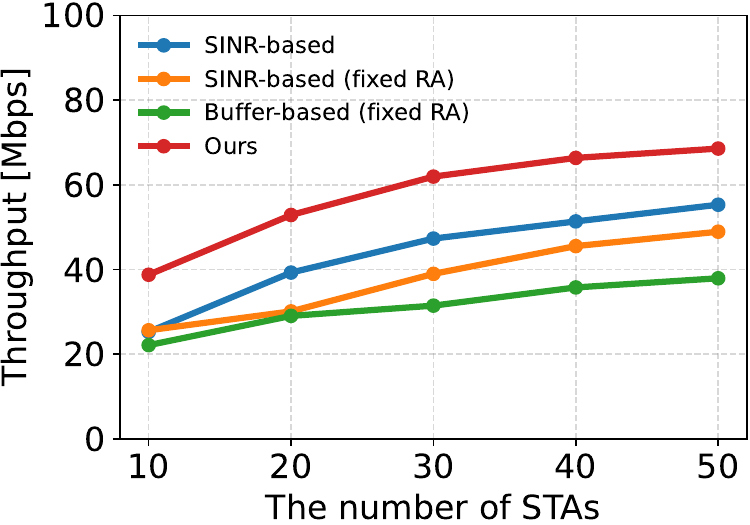}
\label{fig:throughput_STA}
}
\vspace{-6pt}
\caption{{Throughput of the proposed \gls{USRA} algorithm and baseline methods with different packet arrival rates and numbers of STAs, respectively. (a) The number of STAs is 20. (b) The packet arrival rate is set to $200~\text{frames/s}$.}}
\label{fig:throughput}
\vspace{-6pt}
\end{figure}

\begin{figure}[t]
    \centering
    \includegraphics[draft=false, width= \if 1\mycmd 0.5 \else 0.45 \fi \columnwidth]{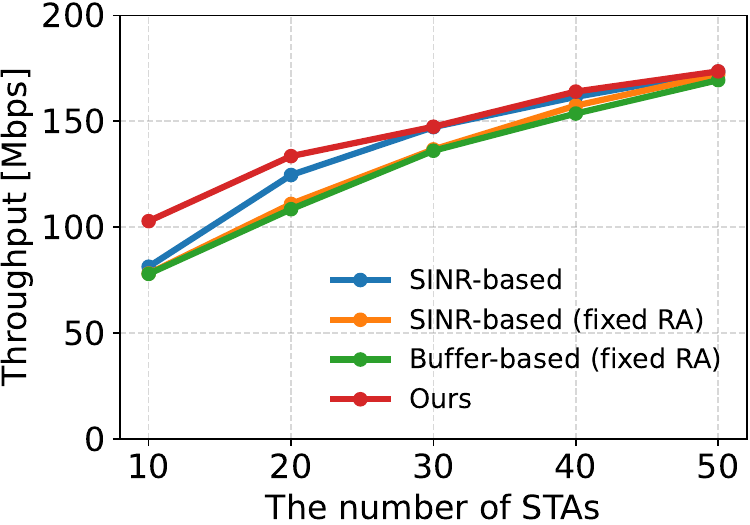}
    \vspace{-8pt}
    \caption{{Throughput of the proposed \gls{USRA} algorithm and baseline methods with various numbers of STAs. The packet arrival rate is set to $10000~\text{frames/s}$.}}
    \label{fig:throughput_saturated}
    \vspace{-5pt}
\end{figure}

\begin{figure}[t]
\centering
\subfigure[]{
\includegraphics[draft=false, width= \if 1\mycmd 0.36 \else 0.45 \fi \columnwidth]{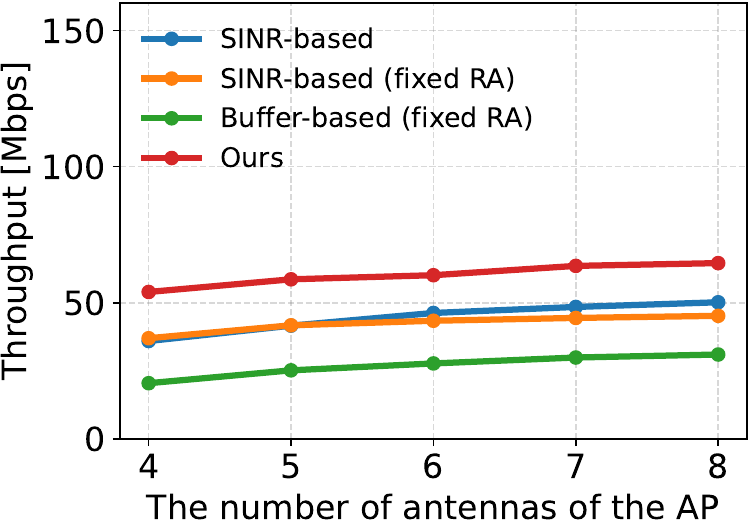}
\label{fig:throughput_ant_1}
}
\subfigure[]{
\includegraphics[draft=false, width= \if 1\mycmd 0.36 \else 0.45 \fi \columnwidth]{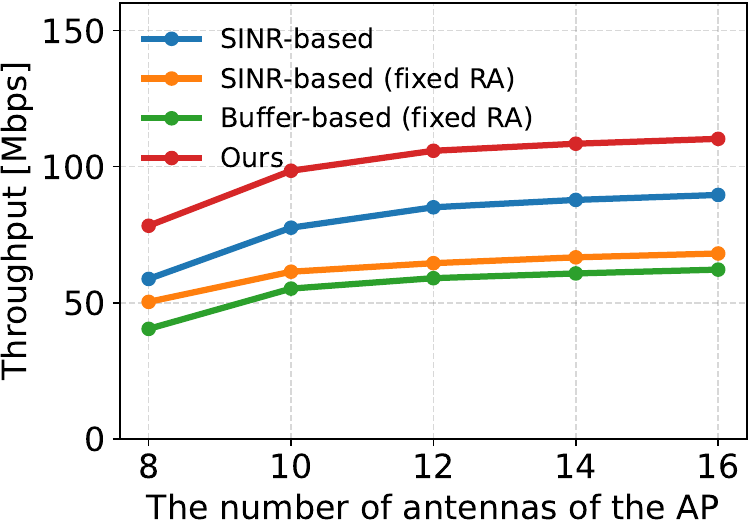}
\label{fig:throughput_ant_2}
}
\vspace{-6pt}
\caption{Throughput across various numbers of antennas for AP and STAs. (a) $N_\text{T} = 1$. (2) $N_\text{T} = 2$. }
\label{fig:throughput_ant}
\vspace{-6pt}
\end{figure}

In Fig. \ref{fig:throughput}, we measure the throughput by varying the number of STAs and packet arrival rates. The SINR-based scheduling algorithm shows higher throughput than both SINR and buffer-based methods with fixed RA in addressing the \gls{USRA} problem, especially under saturated traffic conditions with MU-MIMO and MIMO MS. However, it does not account for the buffer status of STAs. Our method, which includes MU-MIMO, MIMO MS, and STA buffer statuses, outperforms the baselines. At an extreme frame rate of 10000 frames/s, our method achieves throughput levels comparable to or better than the baselines, even in saturated conditions, as shown in Fig. \ref{fig:throughput_saturated}.
In Fig. \ref{fig:throughput_ant}, we assess the throughput as the number of antennas on the AP and STAs increases. Our method consistently surpasses the baseline in all antenna configurations.

\begin{figure}[t]
    \centering
    \includegraphics[draft=false, width= \if 1\mycmd 0.5 \else 0.7 \fi \columnwidth]{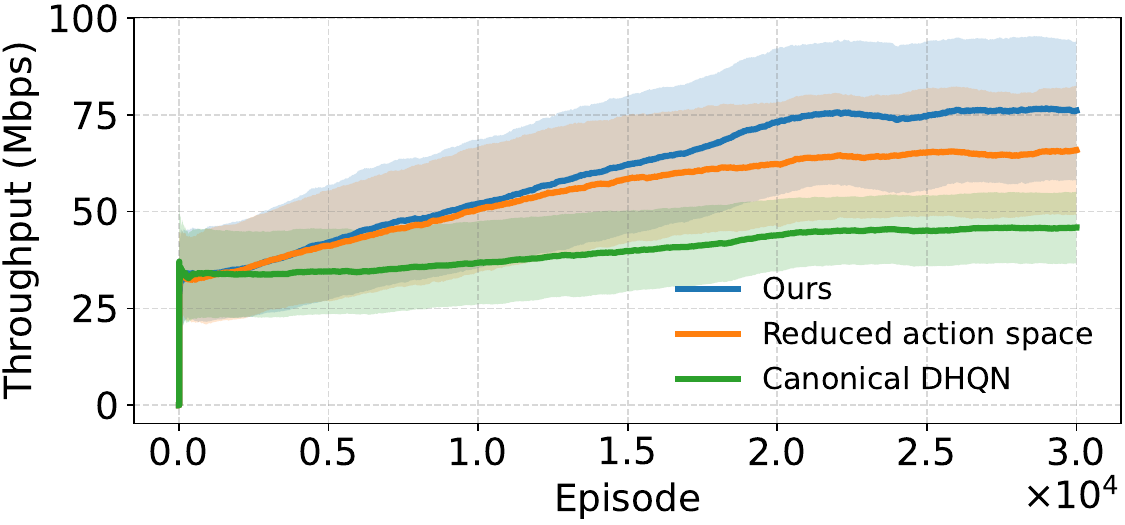}
    \vspace{-8pt}
    \caption{{Throughput of the \gls{USRA} algorithms with canonical DHQN, reduced action space, and proposed methods. The number of STAs is 20, and the packet arrival rate is set to 500 frames/s.}}
    \label{fig:throughput_DQN}
    \vspace{-5pt}
\end{figure}

For the ablation study on sub-agent design for MU-MIMO in Sec. \ref{subagent_design}, we assess the throughput of the DHQN-based \gls{USRA} algorithm with and without reduced action space and channel subspace strategy in Fig. \ref{fig:throughput_DQN}. The standard DHQN, hindered by vast action spaces, only achieves $70~\text{Mbps}$. Although the version with reduced action space shows improved throughput, it struggles with generalizing mutual interference among STAs. In contrast, the optimized DHQN effectively addresses the problem of MU-MIMO user selection.

\section{Conclusion}
\label{sec:conclusion}

This letter proposes a novel DHRL-based \gls{USRA} algorithm, where a master agent selects a combination of RUs from all possible cases, and sub-agents then schedule STAs to the RUs. We optimize the throughput by contemplating STAs' SINRs and buffer status, so that more efficient utilization of time, frequency, and spatial resources is achieved via the joint design of US, RA, and MU-MIMO.
To address the US challenge for MU-MIMO, we propose a sub-agent design with reduced action space and channel subspace strategy.
The numerical results show that the proposed algorithm achieves significantly higher throughput than the existing schemes.

\bibliographystyle{IEEEtran}
\bibliography{{80211ax}}

\end{document}